\documentclass[]{aastex631}

\usepackage{verbatim}
\usepackage{amsmath}
\usepackage{graphicx}
\usepackage{url}
\usepackage{footmisc}

\graphicspath{{./}{figures/}}

\begin{document}

\title{QLP Data Release Notes 003: GPU-based Transit Search}

\correspondingauthor{Michelle Kunimoto}
\email{mkuni@mit.edu}

\author[0000-0001-9269-8060]{Michelle Kunimoto}
\affiliation{Kavli Institute for Astrophysics and Space Research, Massachusetts Institute of Technology, Cambridge, MA 02139}

\author[0000-0002-5308-8603]{Evan Tey}
\affiliation{Kavli Institute for Astrophysics and Space Research, Massachusetts Institute of Technology, Cambridge, MA 02139}

\author[0000-0003-0241-2757]{Willie Fong}
\affiliation{Kavli Institute for Astrophysics and Space Research, Massachusetts Institute of Technology, Cambridge, MA 02139}

\author[0000-0002-2135-9018]{Katharine Hesse}
\affiliation{Kavli Institute for Astrophysics and Space Research, Massachusetts Institute of Technology, Cambridge, MA 02139}

\author{Glen Petitpas}
\affiliation{Kavli Institute for Astrophysics and Space Research, Massachusetts Institute of Technology, Cambridge, MA 02139}

\author[0000-0002-1836-3120]{Avi Shporer}
\affiliation{Kavli Institute for Astrophysics and Space Research, Massachusetts Institute of Technology, Cambridge, MA 02139}

\begin{abstract}
The Quick-Look Pipeline \cite[QLP;][and references therein]{Huang2020,Kunimoto2021} searches for transit signals in the multi-sector light curves of several hundreds of thousand stars observed by TESS every 27.4-day sector. The computational expense of the planet search has grown considerably over time, especially as the TESS observing baseline continues to increase in the second Extended Mission. Starting in Sector 59, QLP has switched to a significantly faster GPU-based transit search capable of searching an entire sector in only $\sim1$ day. We describe its implementation and performance.
\end{abstract}


\section{Introduction}

Since the start of the TESS mission, QLP has used the Box Least Squares (BLS) algorithm \citep{Kovacs2002} implemented in \texttt{VARTOOLS} \citep{HartmanBakos2016} to perform its planet search. Due to the high computational expense, QLP searches an under-sampled frequency grid on the order of 80,000 frequencies or less. Short-period, long-baseline signals are adversely affected as transit signals may not line up well, which hampers post-BLS analysis. QLP also searches for periods up to only $\sim$56 days despite many stars having hundreds of days of data in order to avoid increasing the number of frequencies needed to be searched. Recognizing that searches will be more expensive as the TESS observing baseline increases, QLP now takes advantage of Graphics processing units (GPUs) for a significantly faster algorithm.

\section{Updates}

\subsection{GPU-Based Transit Search}\label{sec:GPU}

The GPU BLS implementation is provided in \texttt{cuvarbase}\footnote{\url{https://github.com/johnh2o2/cuvarbase}}, which uses \texttt{PyCUDA} \citep{pycuda2012} for fast time-series analysis. \texttt{cuvarbase} further optimizes planet searches by using a Keplerian assumption, where only transit durations near the duration expected for a central, circular orbit at a given period and host star density are searched. We adopt stellar density from the TESS Input Catalog \citep{Paegert2021}, and assume a solar density if missing. We search for durations between 0.5 and 2.0 times a circular orbit duration to account for eccentric or grazing orbits.\footnote{We use the following other \texttt{cuvarbase} inputs: \texttt{samples\textunderscore per\textunderscore peak} = 2, \texttt{dlogq} = 0.1, and \texttt{noverlap} = 3.}

Determined for each star, the minimum orbital period searched is the period where the semi-major axis is three times (Sectors 59, 60) or two times (Sectors 61+) the stellar radius, which discounts unphysical orbits and reduces false positive contamination. The maximum period is the length of the longest continuous stretch of data, or half this length if there are no gaps longer than one TESS sector.

\subsection{Transit Statistics}

\texttt{cuvarbase} enables a significantly faster BLS search, but is missing the computation of several useful statistics:

\begin{enumerate}
    \item \textbf{Spectroscopic signal-to-noise:} The basic statistic computed by BLS is the signal residue (SR) as a function of trial frequency $f$. The strongest signal in the light curve will be at the frequency that maximizes the spectroscopic signal-to-noise ratio (S/N),
    \begin{equation}\label{eqn:SN}
        \text{S/N}(f) = \frac{\text{SR}(f) - \bar{\text{SR}}}{\sigma_{\text{SR}}},
    \end{equation}
    where $\bar{\text{SR}}$ and $\sigma_{\text{SR}}$ are the spectrum mean and standard deviation, respectively. SR is outputted by \texttt{cuvarbase}, but we make two changes to Eqn. \ref{eqn:SN}. First, we median-bin the spectrum\footnote{Following Sturge's law, $N = 1 + 3.322 \log{n}$, where $n$ is the number of spectral data points and $N$ is rounded to the nearest integer.} and interpolate over the binned spectrum to find the trend at each point, $\tilde{\text{SR}}(f)$, and remove this instead of $\bar{\text{SR}}$. This is motivated by the fact that BLS spectra feature rising trends toward low frequencies, thus biasing detections toward long periods and outlier/junk-dominated signals. Second, we estimate $\sigma_{\text{SR}}$ as 1.4826 times the median absolute deviation of SR which is more robust to outliers.
    \item \textbf{Trapezoid model fit:} We fit the detected light curve signal using a trapezoid model parameterized by the orbital period ($P$), transit epoch ($T_{0}$), depth ($\delta$), duration divided by orbital period ($q$), duration of ingress divided by transit duration ($q_{\text{in}}$), and out-of-transit magnitude level ($z$). $P$ is fixed to the BLS period while other parameters are fit using initial guesses from BLS. To speed up the fit, we only fit data within two transit durations of the transit center. We place bounds on $q$ between 0 and 1, and $q_{\text{in}}$ between 0 (perfectly box-shaped) and 0.5 (V-shaped).
    \item \textbf{Noise estimates:} Following \texttt{VARTOOLS} calculations, we subtract the trapezoid model from the light curve and estimate the white noise ($\sigma_{w}$) and red noise ($\sigma_{r}$) on the timescale of the transit duration.
    \item  \textbf{Signal-to-pink noise:} \texttt{VARTOOLS} computes the signal-to-pink noise ratio \cite[S/N$_{\text{pink}}$;][]{Pont2006} as
    \begin{equation}\label{eqn:spn}
        \text{S/N}_{\text{pink}} = \frac{\delta}{\sqrt{(\sigma_{w}^{2}/n_{t}) + (\sigma_{r}^{2}/N_{t})}},
    \end{equation}
    where $n_{t}$ is the number of points in transit and $N_{t}$ is the number of transits. However, Eqn.~\ref{eqn:spn} optimistically assumes the transit remains at depth $\delta$ over the entire transit duration, which over-estimates the signal strength of grazing transits. We differentiate from \texttt{VARTOOLS} by replacing $\delta$ with the integral over the full trapezoid model to more realistically take into account transit shape. 
    \item \textbf{Other statistics:} We compute the number of points in, before, and after transit, number of transits, and out-of-transit magnitude level. In order for a transit to count toward the number of transits, we require there to be data within 0.5 transit durations of the expected mid-transit time.
\end{enumerate}

With \texttt{VARTOOLS}, QLP used detection criteria of at least two transits, $\text{S/N}_{\text{pink}} > 9$, and $\text{S/N} > 5$ for stars brighter than $T = 12$ mag and $\text{S/N} > 9$ for fainter stars. We now require $\text{S/N} > 9$ for all stars because S/N values for transit-like signals tend to be higher in the GPU implementation due to the finer frequency grids.

\section{Performance}

We ran both GPU- and CPU-based algorithms on QLP multi-sector light curves from Sector 58 Camera 1 CCD 4 (77567 stars) and Camera 4 CCD 4 (38486 stars). Comparison results are summarized in Table \ref{tab:performance}. Despite searching up to $\sim10$ times more frequencies, the GPU implementation was $\sim40$ times faster. GPU BLS was also able to recover more known TESS Objects of Interest (TOIs). Both algorithms missed dozens of TOIs, but almost all of these had low S/N or periods longer than our search space.

\begin{table}[h]
    \centering
    \begin{tabular}{c|c|c|c}
    \hline\hline
        BLS implementation & Average runtime per star & Average runtime per star & Number of TOIs \\
        & (Camera 1) &  (Camera 4) & recovered \\
    \hline
        GPU-based (\texttt{cuvarbase}) & 1.9 seconds & 4.2 seconds & 392/468 \\
        CPU-based (\texttt{VARTOOLS}) & 77 seconds & 172 seconds & 380/468 \\
    \end{tabular}
    \caption{Comparison in the performance of BLS algorithms, in terms of average runtime per star and the number of TOIs recovered. \texttt{cuvarbase} uses an optimal frequency grid while \texttt{VARTOOLS} used an under-sampled frequency grid due to the time expense.}
    \label{tab:performance}
\end{table}

\section{Acknowledgements}
These data release notes provide processing updates from the Quick-Look Pipeline (QLP) at the TESS Science Office (TSO) at MIT. This work makes use of FFIs calibrated by TESS Image CAlibrator \citep[TICA;][]{TICA}, which are also available as High-Level Science Products (HLSPs) stored on the Mikulski Archive for Space Telescopes (MAST). Funding for the TESS mission is provided by NASA's Science Mission Directorate.

\bibliography{refs}


\end{document}